\documentclass[10pt]{article}
\usepackage{graphicx,color}
\usepackage{makeidx}
\usepackage{float}
\usepackage{setspace}
\usepackage[section]{placeins}
\usepackage{amsmath}
\usepackage{subfig}
\usepackage{framed}
\def\qed{\rule[0mm]{.6em}{.6em}}
\begin{document}

\bibliographystyle{plain}
\setcounter{secnumdepth}{3}
\setcounter{tocdepth}{3}
\newtheorem{Lemma}{Lemma}[section]
\newtheorem{theorem}{theorem}[section]
\newtheorem{Corollary}{Corollary}[section]
\newenvironment{Proof}{{\bf Proof }}{\qed}
\newenvironment{Algorithm}{\medskip\small\mbox{}\\\begin{tabbing}\tt}{\end{tabbing}\normalsize}
\newenvironment{Subheading}{\medskip\mbox{}\\ \noindent \bf}{\rm \smallskip\\}
\newcommand{\len}{{\tt{Len}}}
\newcommand{\str}{{\tt{Str}}}
\newcommand{\II}{{\cal I}}
\newcommand{\MM}{{\cal M}}
\newcommand{\DD}{{\cal D}}
\newcommand{\CC}{{\cal C}}
\newcommand{\BB}{{\cal B}}
\newcommand{\KK}{{\cal K}}
\newcommand{\WW}{{\cal W}}
\newcommand{\XX}{{\cal X}}
\newcommand{\YY}{{\cal Y}}
\newcommand{\LL}{{\cal L}}
\newcommand{\ZZ}{{\cal Z}}
\newcommand{\UU}{{\cal U}}
\newcommand{\VV}{{\cal V}}
\newcommand{\TT}{{\cal T}}
\newcommand{\SSS}{{\cal S}}
\newcommand{\RR}{{\cal R}}
\newcommand{\HH}{{\cal H}}
\newcommand{\CA}{{\cal A}}
\newcommand{\PP}{{\cal P}}
\newcommand{\FF}{{\cal F}}
\newcommand{\ZM}{{\cal M}}
\newcommand{\F}{{\cal F}}
\newcommand{\Q}{{\cal P}}
\newcommand{\N}{{\cal N}}
\newcommand{\lra}{\leftrightarrow}
\newcommand{\la}{\leftarrow}
\newcommand{\ua}{\uparrow}
\newcommand{\nwa}{\nwarrow}
\newcommand{\da}{\nwarrow}

\newcommand{\nn}{\noindent}
\newcommand{\mm}{\medskip}
%\newcommand{\ds}{\displaystyle}
%\let\oldfootnote\footnote
%\renewcommand{\footnote}[1]{\oldfootnote{\doublespacing #1}}
%\let\oldcaption\caption
%\newcommand{\Caption}[1]{\caption{\small #1}}
%\renewcommand{\Caption}[1]{\caption{\small \doublespacing #1}}

%\input{perpaper}
%\input{perACM}

%\documentclass[12pt]{article}
%\begin{document}
%\bibstyle{plain}

\begin{center}

{\bf \Large Persistent Phylogeny: A Galled-Tree and Integer Linear Programming Approach}

\vspace{7mm}
Dan Gusfield\footnote{Computer Science, University of California, Davis} 
\end{center}

\begin{abstract}
The Persistent-Phylogeny Model is an extension of the widely studied Perfect-Phylogeny Model,
encompassing a broader range of evolutionary phenomena. Biological and algorithmic questions
concerning persistent phylogeny have been intensely investigated in recent years. In this paper,
we explore two alternative approaches to the persistent-phylogeny problem that grow out of our previous
work on perfect phylogeny, and on galled trees. 
We develop an integer programming solution to the Persistent-Phylogeny Problem; empirically explore its
efficiency; and empirically explore the utility of using fast algorithms that recognize galled trees, to recognize persistent
phylogeny. 
The empirical results identify parameter ranges where persistent phylogeny are galled trees with high frequency, and
show that the integer programming approach can efficiently identify persistent phylogeny of much larger size than has
been previously reported.
\end{abstract}

\section{The Perfect Phylogeny Problem for Binary Characters}

The Persistent-Phylogeny Model is an extension of the Perfect-Phylogeny Model, so we begin with a brief discussion
of perfect phylogeny. 

\medskip

\noindent{\bf Definition} Let $M$ be an $n$ by $m$ matrix representing 
$n$ taxa in terms of $m$ characters or traits that describe the 
taxa.  Each character takes on one of two possible {\it states}, 0 or 1; 
a cell $(f,c)$ of $M$ has a value of one if and only if the state of character $c$ is 1 
for taxon $f$.
Thus the characters of $M$ are {\it binary-characters} and $M$ is called
a {\it binary matrix}. 

%When a taxon $f$ has state 1 for a binary character $c$, we also say that ``$f$ possesses (or has) character $c$." 

\noindent {\bf Definition} Given an $n$ by $m$ binary-character matrix $M$ for $n$ 
taxa, a {\it perfect phylogeny for $M$} with {\it all-zero} root sequence is a {\it rooted} (directed) tree $T$ with exactly 
$n$ leaves, obeying the following properties: 

\begin{enumerate}
\item
Each of the $n$ taxa labels exactly one leaf of $T$. 

\item
Each of the $m$ characters labels {\it exactly one} edge of $T$. 

\item
For any taxon $f$, the characters that label the edges along the unique path 
from the root to the leaf labeled $f$, specify all of the characters that taxon $f$ 
possesses (i.e., whose state is 1). 

\end{enumerate} 

The key biological assumption that leads to the perfect phylogeny model is that
in the evolutionary history of the taxa, each character mutates from the zero state 
to the one state {\it 
exactly} once, and {\it never} from the one state back to the zero state. Hence every character $c$ labels
exactly one edge $e$ in a perfect phylogeny $T$ for $M$, indicating
the unique point in the evolutionary history of the taxa when 
character $c$ mutates. A character that has this property is called a {\it perfect character}
\cite{CG2012,Hem2010,Hillis1999,RXSB2006,RH00}. 
%In population genetics, perfect characters are motivated by the {\it infinite-sites} 
%model \cite{HK85} and widely collected {\it single nucleotide polymorphism (SNP)} data (for example, see \cite{HapMap3}). 
Of course,
most evolutionary characters are not perfect, but perfect (or near-perfect) characters are sufficiently frequent to motivate the
study of perfect phylogeny.
See \cite{Recbook2014} for a detailed discussion of
the interpretation of perfect phylogenies, and of biological settings where perfect phylogenies are observed or
hypothesized to exist. Additional recent examples of perfect characters and perfect phylogenies 
(not discussed in \cite{Recbook2014}) come from single-cell studies of mutating tumors (as one of several recent examples, see \cite{GUND2015}).

\medskip

{\bf The perfect-phylogeny problem:}  Given an $n$ by $m$ binary matrix $M$,
determine whether there is a perfect phylogeny for $M$, and if so, build
one.

The perfect-phylogeny problem can be solved in polynomial (even linear) time. 
The following theorem is well-known and explained in many places, for example see \cite{GUS97,Recbook2014}.

\begin{theorem}{\bf The perfect-phylogeny theorem} 
\label{tperfectphylogeny}
Matrix $M$ has a perfect phylogeny (with all-zero ancestral sequence) 
if and only if no pair
of its columns $c,d$ contains the {\it three} binary pairs 0,1; 1,0; and 1,1.
\end{theorem}

Any pair of columns that contain all three binary pairs are called {\it conflicted} columns, and a column that
is not conflicted with any other column is called {\it unconflicted}.
%By direct appliction of Theorem \ref{tperfectphylogeny}, the perfect-phylogeny problem can be solved in $O(nm^2)$,
%where each comparison operation and each reference to $M$ takes one time unit.  
%A more efficient solution is possible; the problem can be solved in  $O(nm)$ time \cite{G91}.

\subsection{Dollo Parsimony and Persistent Phylogeny}

Several extensions of the perfect character model have been proposed in order to address a wider range of
evolutionary phenomena. In the {\it Dollo} (Parsimony) model \cite{Dollo,FEL04,KooninDollo2006}, 
each character can mutate from the 0 state to the 1 state
at most once in the history (as in the perfect phylogeny model), but the character can mutate from the 1 state
back to the 0 state at any point where the character has state 1. This models evolutionary characters that
are gained with low probability, but that
are lost with much higher probability (allowing the 1 to 0 mutation without constraint).
The Dollo model is appropriate
``for reconstructing evolution of the gene repertoire of eukaryotic organisms because although
multiple, independent losses of a gene in different lineages are common, multiple gains of the same
gene are improbable \cite{KooninDollo2006}."

More recently, a more limited version of the Dollo  model was proposed, where any character can mutate from 
state 0 to state 1 at most once in the history, and symmetrically, it can mutate from state 1 to state 0
at most once. This model is called the {\it Persistent-Phylogeny} model, or the {\it Persistent-Perfect-Phylogeny} model. It
models evolutionary characters that are gained with low probability, and then are lost with low (but not-zero) probability.
%We will use the first term in order to avoid confusion with the Perfect-Phylogeny model.  
%When a set of data, $M$, can
%be generated on a tree that obyes the assumptions of the persistent-phylogeny model, we say that $M$ can be represented
%by a persistent phylogeny.  
The Persistent-Phylogeny model was first 
proposed and further examined in the papers 
\cite{Przyt07,ZhengPrzyt07,Przyt07stable,BBDT12,BCDDP2013,BC2014arX,BC2014}.

\medskip

{\bf The Persistent-Phylogeny Problem:}  Given an $n$ by $m$ binary matrix $M$,
determine whether $M$ can be represented by a persistent phylogeny for $M$, and if so, build
one.
\medskip

\subsection{Persistent Phylogeny and Galled Trees}

The complexity of the persistent-phylogeny problem is open -- there is no known polynomial-time algorithm for the problem, and neither has it been shown to be NP-complete.
Of course, if $M$ has a perfect phylogeny, then it has a persistent phylogeny, so
a polynomial-time special case of the persistent phylogeny problem 
is the case of data that has a perfect phylogeny. A more interesting polynomial-time special case 
is that of a {\it galled tree}, which might be a directed tree, but
might be a  particular type of directed acyclic graph (DAG). In detail, a galled tree is a rooted DAG $G$,
where all cycles in the underlying undirected graph of $G$ are {\it node disjoint}.  Trivially, every perfect phylogeny
is a galled tree, but many datasets that cannot be represented on a perfect phylogeny can be represented
on a galled tree.
A full, formal definition of a galled tree can be found in \cite{GELOPT04,Recbook2014}. The 
relationship between persistent phylogeny and galled trees was developed in \cite{GELOPT04} and also discussed
in \cite{Recbook2014}. That relationship is summarized as follows:

\begin{theorem} 
If binary data $M$ can be represented by a galled tree, then $M$ can be represented by a persistent phylogeny.
Moreover, the galled tree for $M$ can be converted to a persistent phylogeny for $M$ in linear time.
\end{theorem}

The question of whether a binary data $M$ can be represented by a galled tree has a polynomial-time 
solution \cite{GELOPT04,Recbook2014},
and a practical implementation as program {\it galledtree.pl}, 
which is available through this authors website \cite{GUSWEB}. The program is very fast, and (as detailed later in the paper) solved
every problem instance examined for this paper in under one second.
Thus, any approach to the general persistent phylogeny
problem needs only concentrate on data that is not representable by a galled tree. This naturally leads to the
question of {\it how frequently} data that is representable on a persistent phylogeny is also representable by a galled tree.
This paper, in part, addresses that question through empirical testing. The results are discussed in section \ref{sempirical}.

\section{Solving the Persistent-Phylogeny Problem by Integer Programming}

When data $M$ is not representable by a galled tree, it might still be representable by a persistent phylogeny, and so
we would like a practical method to solve instances of the persistent-phylogeny problem on problem sizes as large as possible. 
Efforts to develop
such algorithms, and test their efficacy,  appear in \cite{BBDT12,BCDDP2013,BC2014}.
Here we develop and study a practical method that can solve the persistent-phylogeny problem using {\it integer linear programming}, on
instances of larger size than have previously been reported.
Integer linear programming has been successful in efficiently solving many hard problems on instances whose size and structure is of
relevance in current applied domains.

\section{An ILP Solution to Persistent-Phylogeny Problem}

The persistent-phylogeny problem was shown in \cite{BBDT12} to be reducible to a problem called the {\it Incomplete perfect-phylogeny
with persistent completion (IP-PP)} problem. The integer programming solution in this paper follows that approach, solving  instances of
the persistent phylogeny problem by solving instances of the IP-PP problem. Next, we define the IP-PP problem,
and its integer programming formulation.

\medskip

\noindent {\bf Definition} \cite{BBDT12}:  Given a binary matrix $M$, the {\it extended matrix} $M_e$ contains two columns, $j_1$ and $j_2$,  
for each column $j$ in $M$. Column $j_1$ of $M_e$ is derived from column $j$ in $M$ 
by replacing every occurrence of `0' in column $j$ of $M$ with `?' in column $j_1$ of $M_e$.
Column $j_2$ of $M_e$ is derived from column $j_1$ by replacing every occurrence of `1' in $j_1$ with `0'. 
See Figure \ref{fpersistent}.

\begin{figure}
$M~~~~~~~~~~M_e~~~~~~~M'_e = \verb|Completion of | M_e$\\
1110~~~~~     101010??~~~~~    10101000\\
0111~~~~~     ??101010~~~~~    11101010\\
0000~~~~~     ????????~~~~~    00000000\\
1010~~~~~     10??10??~~~~~    10001000\\
1100~~~~~     1010????~~~~~    10101100\\
1111~~~~~     10101010~~~~~    10101010\\
\caption{M, $M_e$ and $M'_e$}
\label{fpersistent}
\end{figure}

Note that for any pair of columns $(j_1, j_2)$ in $M_e$ that are derived from the same column $j$ in $M$, 
column $j_1$ contains a '?' in a row $r$ 
if and only if column $j_2$ also contains a '?' in row $r$. 
We call such cells {\it twin cells}. 

\medskip

\noindent {\bf Definition} \cite{BBDT12}: A {\it completion} $M_e'$ of an extended matrix $M_e$ derived from binary matrix $M$, 
is obtained by replacing each '?' in $M_e$
with '0' or '1', subject to the constraint that for any column $j$ in $M$, if $M_e(r,j_1) =~ ?$, then $M_e'(r,j_1) = M_e'(r,j_2)$.
That is, the values given to twin cells must be the same.
See Figure \ref{fpersistent}. 

The following theorem, stated and proved in \cite{BBDT12}, is central to the integer programming approach developed in this paper.

\begin{theorem}
\label{tcompletion}
Let $M_e$ be the extended matrix obtained from binary matrix $M$. Then $M$ 
can be represented by a persistent phylogeny if and only if there is a completion $M_e'$ of
$M_e$ such that $M_e'$ can be represented by a perfect phylogeny.
\end{theorem}

Given $M$, the {\it IP-PP} problem is the problem of determining if there is a completion of $M_e$ that can be represented by a 
perfect phylogeny.

\subsection{The Integer Linear Program for the IP-PP problem}

The integer linear programming approach to solving the IP-PP problem is an extension of the ILP formulation for the
{\it Incomplete directed perfect phylogeny (IDPP)} problem which is defined next.

\begin{quote}
Given an $n$ by $m$ binary matrix $M$, with a set of cells $\KK$ that have missing values,  find \{0,1\} values 
to assign to the cells in $\KK$ so that the resulting matrix $M'$ has a perfect phylogeny with all-zero ancestral sequence;
or determine that there is no such assignment. When there is such an assignment, we call it a {\it solution} to the IDPP problem.
\end{quote}

The IDPP problem actually has a polynomial-time solution \cite{PPSS}, but we do not make use of it here. Rather, we use an ILP
approach to a variant of IDPP problem from \cite{GFB07}. In that ILP approach, 
there is one binary variable $Y(i,j)$ for each cell $(i,j)$ in $\KK$.
The core of the ILP formulation specifies linear inequalities that constrain the values given to the $Y$ variables 
so that the resulting matrix $M'$ satisfies Theorem \ref{tperfectphylogeny}, i.e., the necessary and sufficient conditions
for the data to be representable by a perfect phylogeny. Consequently,  for a dataset $M$, the IDPP problem has a solution
if and only if the corresponding ILP instance is feasible.

We can easily modify the ILP formulation for the IDPP problem to obtain an ILP solution to the IP-PP problem. In particular, given
matrix $M$, we build the extended matrix $M_e$ from $M$, and consider each cell with a `?' to be a cell with a missing value.
Then, we construct the ILP formulation for the IDPP problem for input matrix $M_e$, with the added equality ``
$Y(r,j_1) = Y(r,j_2)$" for each pair of twin cells, $(r,j_1), (r,j_2)$. 

These equalities  assure that twin cells receive the same values.
We call the resulting ILP formulation, modified from an IDPP formulation, the {\it MIDPP} formulation for input $M$.
We implemented the MIDPP formulation by extending the previously developed software for the IDPP
problem, available on the author's webpage \cite{GUSWEB}. The software (called {\it PERILP.pl}, written in Perl) 
that creates an MIDPP formulation given $M$
is also available there.

\subsubsection*{In review}
The acronyms are similar and confusing, so here we summarize the integer linear programming  solution to the persistent phylogeny problem.
Binary data $M$  can be represented by a persistent phylogeny, if and only if the extended matrix $M_e$ has
a completion $M'_e$ that can be represented by a perfect phylogeny; if and only if the MIDPP integer programming formulation for $M_e$
has a feasible solution. 
%Equivalently, the Persistent Phylogeny problem reduces
%to the  IP-PP problem, which reduces to the MIDPP (ILP) formulation, which can be solved by an ILP solver.

\section{Empirical Evaluation}
\label{sempirical}
We conducted extensive empirical testing,
under two approaches to data generation, and differing combinations of parameters, 
to answer several questions:

\begin{quote}
1) How frequently
is data that is representable by a persistent phylogeny also representable by a galled tree?

2) When data is representable by a galled tree, how quickly can a galled tree be found by
specialized galled tree software, and how quickly can a persistent phylogeny by found for that
data by the ILP approach for the general persistent phylogeny problem?

3) When data is representable by  a persistent phylogeny, but not a galled tree, how quickly can 
a persistent phylogeny for that data be found by the ILP approach?

4) When data is not representable by a persistent phylogeny, how quickly can the ILP method determine this?
\end{quote}

Our empirical tests varied $n$, the number of rows; $m$, the number of columns; and $bp$, the probability of a back-mutation
occurring on an edge. 
As detailed in the tables below, $n$ ranged from 40 to 1000, and $m$ ranged from 30 to 500, and $bp$ was selected from
the set $\{0.01, 0.05, 0.2, 0.4\}$.
For each combination of parameters, 50 individual datasets were generated, except when $n = 1000$, where only 25 datasets
were generated.

\subsection{Data Generation}

Data was generated in two ways, one that guaranteed data that can be represented by a persistent phylogeny,
and one that only guaranteed that the data can be represented by a phylogeny under the Dollo model. 
However, in the second case, the
generated data only infrequently lacked a persistent phylogeny.
Thus, the empirical testing was mostly directed at data that could be represented by a persistent phylogeny.

Here we first describe how data for a single problem instance is generated when the data is {\it guaranteed} to be represented
by a persistent phylogeny.
To start, the program MS \cite{HU2002} is used to generate data, denoted $D$, with a specified number of
rows ($n$) and a specified number of columns ($m$) that satisfies the perfect phylogeny condition given in 
Theorem \ref{tperfectphylogeny}. 
%In program MS, this is accomplished by setting  the recombination parameter to 0.  
Let $T$ denote the perfect phylogeny that MS generates for $D$.
Then, conceptually, but with a faster implementation,
the algorithm successively walks from the root of $T$ to each leaf, to determine where
any back mutations will occur.  In detail, 
when an edge $e$ is traversed during the walk of $T$, the program
generates a random number $r$; and if $r < bp$, 
the program tries to find a character to back mutate at $e$. It does this by randomly choosing a character $c'$ 
that was mutated to state one on an edge $e'$ leading to $e$, such that character $c'$ has not yet been
back-mutated anywhere in $T$. If such a character $c'$ is found, then $c'$ is back-mutated on edge $e$. 
The modified tree $T$ is thus a persistent phylogeny for a dataset $M$, i.e., the sequences at the leaves of $T$, generated
on the walks in $T$.
Note that despite being generated on a unconstrained persistent phylogeny
with back mutations, $M$ might still be representable by a galled tree, or a perfect phylogeny.

The same approach is followed when generating data for the Dollo model, but if $r < bp$ at edge $e$, we only
require that character $c$ has not been back-mutated on any edge leading to $e$.
Again, although the data is generated
under the unconstrained Dollo model, it is possible that the
data generated is representable on a persistent phylogeny, a galled tree, or even a perfect phylogeny.

Once $M$ is generated, all {\it unconflicted} columns are removed from $M$, since this makes the problem instance smaller, and it
is known that such removals do not affect the existence or non-existence of persistent phylogenies. In fact, this
is a consequence of a more general fact about unconflicted columns (\cite{Recbook2014}).
Data generation and the ILP solution can also be easily extended to the case when no ancestral sequence is known,
but that discussion is omitted here.

%\subsubsection{Implementation Speedup}

%We note that we do not need to explicitly construct tree $T$ to create $M$ from $D$. Instead, we 
%sort the columns of $D$ in order of the number of ones that appear in the column, 
%largest first.
%This guarantees that for any taxon $t$, the characters that $t$ possesses are sorted, left to right, in the order that they
%appear on the walk from the root of $T$ to the leaf labeled by $t$. Hence, to implement that walk, we simply scan row $t$ of the
%sorted $D$ (left to right), noting when encountering a cell with value 1, and entering the associated character into a list $L$. 
%Each character $c$ in $L$ represents the edge $e$ in $T$ where
%character $c$ mutates. When encountering a cell with value 1, representing character $c$,  we generate a random number $r$ between
%0 and 1, and if $r < bp$, we randomly select a character
%$c' \neq c$ from $L$, provided that $c'$ has not previously been back-mutated.
%

\subsection{Empirical Results}

All programs used to generate the test data and create the ILP formulations, and analyze the results, were written in Perl. 
The ILP solver we used was Gurobi 6.0, running on
a 2.3 GHz Macbook Pro with intel Core i7 (four cores, and up to eight threads). The macbook ran OS X version 10.9.5.  
In our trials, whenever the feasible solution to the ILP was found (so by Theorem \ref{tcompletion}, 
there should be a persistent phylogeny for
the dataset $M$), the values of the $Y(i,j)$ variables were used to form a completion, $M'_e$, of $M_e$, and to build
a perfect phylogeny for $M'_e$, and a persistent phylogeny for $M$, verifying constructively that the programs ran correctly.
All of the programs written in Perl are available  on the author's webpage, and
Gurobi is free to academics and researchers. Thus, all the results presented here can be independently verified (or contradicted - 
we hope not), by the readers. 

Results for data generated on a persistent phylogeny, are shown in 
tables \ref{t2} and \ref{t4}. Galled tree computations were run only on
the datasets in the first table, due to size limitation on the galled-tree program.
Each line in a table shows the results for a particular combination of parameters. 
For each combination of parameters, fifty datasets were generated and analyzed.
The first column in the table shows the number 
of rows ({\it r}) and columns ({\it c}) in the dataset.  The second column ({\it br}) shows the back-mutation rate used to generate the
data (explained in the text).
When galled-tree computations were run, the third column ({\it Gtime}) shows the total reported time (two digits after the decimal point) 
to test the fifty datasets to see which of them
can be represented by a galled tree. Thus, only a fraction of a second is needed per dataset to check if it can be represented on
a galled tree. The fourth column ({\it data types}) shows the 
number of datasets that can
be represented on a perfect phylogeny ({\it perf}); or on a galled tree ({\it gt}), but not a perfect phylogeny; or on 
a persistent phylogeny ({\it pers}), but not a galled tree.  Because of the way these data were generated, those numbers add to the number of datasets
tested for that combination of parameters (50 in this table). The fifth column ({\it conflicts}) shows the average number of
pairwise conflicts observed in the datasets; these numbers are reported for each of the three types of datasets 
in the same order as in column four. 
The sixth column ({\it ILP-data}) gives information about the performance of the ILP solver on the data. The first number ({\it inf}) is
the number of datasets that were determined to be  {\it infeasible}  by the ILP solver, meaning that the data could not
be represented on a persistent phylogeny. By the way these data were generated, that number should be zero in these tables,
and so that entry is only used as a consistency check. The second number in ILP-data 
is the number of ILP computations that were interrupted ({\it int})
because they exceeded the {\it six-minute} time limit. The third number ({\it tm-gt}) is the average time taken by the ILP solver on those
datasets that (independently) were determined to be representable by a galled tree ({\it gt});  none of these were interrupted.
The fourth number ({\it tm-pers}) in ILP-data is the average time
taken by the ILP solver on those datasets that can be represented by a persistent phylogeny ({\it pers}), but not a galled tree, and 
where the ILP execution was not interrupted.

\paragraph*{Datasets not guaranteed to have a persistent phylogeny}
As discussed earlier, we also generated datasets that might not be representable by a persistent phylogeny,
but could be generated on a tree under the Dollo model. About \%10 of the generated data were not
representable on a persistent phylogeny. The running times of these data were consistent with the
datasets reported earlier, and we omit an explicit discussion here due to space limitations.

\subsection{The most striking results}
The most striking result is how efficiently the ILP approach solves the persistent phylogeny problem, with the
examined data. In data generated from a persistent phylogeny, the majority of the datasets were solved in under one second,
and most solved in a handful of seconds.  The six-minute time limit was never reached until 
datasets of 400 taxa and 400 sites.
The results also verify that the running times are highly sensitive to the number of sites, and less sensitive
to the number of taxa and the number of conflicts in the data. These qualitative observations were 
first reported in \cite{BBDT12,BC2014}.
The sizes of the datasets, the complexities of the dataset (measured in the number of conflicting pairs),  
and the times reported, compare very favorably to the
best unconstrained results reported earlier in the literature (for example, see Tables numbered 1 in \cite{BBDT12} and
\cite{BC2014}).

The second most significant result is that many of the datasets that are representable by a 
persistent phylogeny are also representable by a galled tree. Since the time needed to determine if a dataset $M$
is representable by a galled tree is typically much less than the time needed to determine if $M$ is
representable by a persistent phylogeny, the empirical results in this paper suggest that when trying to 
determine if a dataset is representable by a persistent phylogeny, one should first 
determine if $M$ can be represented by a galled tree.

\begin{table}[ht]
\begin{center}
\begin{tabular}{cccccc}  

r;~~c & br & Gtime & data types & conflicts &  ILP-data\\  
       &    &       &perf;~gt;~pers&gt;~pers&inf;~int;~tm-gt;~tm-pers\\
\hline 
\hline 
40,~ 30 & 0.02 & 0 & 39,~10,~ 1 & 2.29,~ 3 & 0,~0,~ 0,~  0 \\ 
40,~ 30 & 0.05 & 1 & 29,~17,~ 4 & 1.7,~ 3.75 & 0,~0,~ 0,~  0 \\ 
40,~ 30 & 0.1 & 0 & 20,~23,~ 7 & 1.73,~ 5 & 0,~0,~ 0,~  0.01 \\ 
40,~ 30 & 0.2 & 0 & 7,~21,~ 22 & 1.95,~ 6.13 & 0,~0,~ 0,~  0 \\ 
\hline 
40,~ 100 & 0.02 & 3 & 12,~17,~ 21 & 7.35,~ 12.85 & 0,~0,~ 0.01,~  0.04 \\ 
40,~ 100 & 0.05 & 3 & 3,~6,~ 41 & 4.66,~ 24.56 & 0,~0,~ 0,~  0.16 \\ 
40,~ 100 & 0.1 & 3 & 0,~0,~ 50 & 0,~ 33.15 & 0,~0,~ 0,~  0.34 \\ 
40,~ 100 & 0.2 & 3 & 0,~0,~ 50 & 0,~ 48.7 & 0,~0,~ 0,~  0.62 \\ 
\hline 
\hline 
60,~ 30 & 0.02 & 0 & 39,~10,~ 1 & 1.8,~ 3 & 0,~0,~ 0,~  0 \\ 
60,~ 30 & 0.05 & 0 & 26,~17,~ 7 & 2.11,~ 4.28 & 0,~0,~ 0,~  0 \\ 
60,~ 30 & 0.1 & 0 & 21,~22,~ 7 & 1.63,~ 4.71 & 0,~0,~ 0,~  0 \\ 
60,~ 30 & 0.2 & 0 & 12,~11,~ 27 & 1.81,~ 5.03 & 0,~0,~ 0,~  0.01 \\ 
\hline 
60,~ 100 & 0.02 & 4 & 11,~23,~ 16 & 4.56,~ 13.56 & 0,~0,~ 0,~  0.06 \\ 
60,~ 100 & 0.05 & 3 & 1,~6,~ 43 & 6,~ 18.83 & 0,~0,~ 0.01,~  0.13 \\ 
60,~ 100 & 0.1 & 4 & 0,~2,~ 48 & 5.5,~ 36.29 & 0,~0,~ 0.03,~  0.47 \\ 
60,~ 100 & 0.2 & 4 & 0,~1,~ 49 & 12,~ 56.04 & 0,~0,~ 0.1,~  0.95 \\ 
\hline 
\hline 
100,~ 30 & 0.02 & 0 & 41,~8,~ 1 & 2.62,~ 5 & 0,~0,~ 0,~  0 \\ 
100,~ 30 & 0.05 & 1 & 33,~16,~ 1 & 1.56,~ 2 & 0,~0,~ 0,~  0 \\ 
100,~ 30 & 0.1 & 1 & 18,~20,~ 12 & 1.45,~ 4.75 & 0,~0,~ 0,~  0 \\ 
100,~ 30 & 0.2 & 1 & 14,~21,~ 15 & 2.14,~ 6 & 0,~0,~ 0,~  0.01 \\ 
\hline 
100,~ 60 & 0.02 & 2 & 15,~20,~ 15 & 2.8,~ 6.46 & 0,~0,~ 0,~  0.01 \\ 
100,~ 60 & 0.05 & 2 & 10,~16,~ 24 & 2.31,~ 11.5 & 0,~0,~ 0,~  0.03 \\ 
100,~ 60 & 0.1 & 1 & 1,~8,~ 41 & 2.37,~ 11.92 & 0,~0,~ 0,~  0.04 \\ 
100,~ 60 & 0.2 & 2 & 0,~6,~ 44 & 3.16,~ 15.75 & 0,~0,~ 0,~  0.11 \\ 
\hline 
100,~ 100 & 0.02 & 4 & 8,~16,~ 26 & 4.12,~ 15.96 & 0,~0,~ 0,~  0.07 \\ 
100,~ 100 & 0.05 & 5 & 2,~9,~ 39 & 5.11,~ 19.17 & 0,~0,~ 0.01,~  0.13 \\ 
100,~ 100 & 0.1 & 5 & 0,~0,~ 50 & 0,~ 34.24 & 0,~0,~ 0,~  0.46 \\ 
100,~ 100 & 0.2 & 5 & 0,~0,~ 50 & 0,~ 44.08 & 0,~0,~ 0,~  1.22 \\ 
\hline 
\hline150,~ 80 & 0.02 & 3 & 13,~37,~ 0 & 1.48,~ 0 & 0,~0,~ 0.09,~  0 \\ 
150,~ 80 & 0.05 & 4 & 0,~0,~ 50 & 0,~ 13.84 & 0,~0,~ 0,~  0.41 \\ 
150,~ 80 & 0.1 & 3 & 0,~0,~ 50 & 0,~ 18.82 & 0,~0,~ 0,~  0.52 \\ 
150,~ 80 & 0.2 & 4 & 1,~0,~ 49 & 0,~ 31.69 & 0,~0,~ 0,~  1.54 \\ 
\hline 
150,~ 100 & 0.02 & 5 & 9,~0,~ 41 & 0,~ 7.41 & 0,~0,~ 0,~  0.06 \\ 
150,~ 100 & 0.05 & 6 & 2,~0,~ 48 & 0,~ 21.12 & 0,~0,~ 0,~  0.25 \\ 
150,~ 100 & 0.1 & 6 & 0,~0,~ 50 & 0,~ 33.74 & 0,~0,~ 0,~  2.2 \\ 
150,~ 100 & 0.2 & 6 & 5,~0,~ 45 & 0,~ 56.04 & 0,~0,~ 0,~  3.67 \\ 
\end{tabular}
\caption{Table for 40, 60, 100 and 150 taxa representable by a persistent phylogeny.}
\label{t2}
\end{center}
\end{table}

\begin{table}[ht]
\begin{center}
\begin{tabular}{cccccc}  
r;~~c & br & Gtime & data types & conflicts &  ILP-data\\  
       &    &       &perf;~gt;~pers&gt;~pers&inf;~int;~tm-gt;~tm-pers\\
\hline 
\hline 
200,~ 100 & 0.02 & -- & 6,~0,~ 44 & 0,~ 11.45 & 0,~0,~ 0,~  0.04 \\ 
200,~ 100 & 0.05 & -- & 1,~0,~ 49 & 0,~ 21.69 & 0,~0,~ 0,~  0.22 \\ 
200,~ 100 & 0.1 & -- & 0,~0,~ 50 & 0,~ 34.86 & 0,~0,~ 0,~  0.64 \\ 
200,~ 100 & 0.2 & -- & 0,~0,~ 50 & 0,~ 44.34 & 0,~0,~ 0,~  0.93 \\ 
\hline 
200,~ 200 & 0.02 & -- & 0,~0,~ 50 & 0,~ 54.9 & 0,~0,~ 0,~  1.33 \\ 
200,~ 200 & 0.05 & -- & 0,~0,~ 50 & 0,~ 101.78 & 0,~0,~ 0,~  3.67 \\ 
200,~ 200 & 0.1 & -- & 0,~0,~ 50 & 0,~ 138.56 & 0,~0,~ 0,~  9.42 \\ 
200,~ 200 & 0.2 & -- & 0,~0,~ 50 & 0,~ 173.78 & 0,~0,~ 0,~  13.67 \\ 
\hline
200,~ 250 & 0.02 & -- & 0,~0,~ 50 & 0,~ 86.86 & 0,~0,~ 0,~  4.01 \\
200,~ 250 & 0.05 & -- & 0,~0,~ 50 & 0,~ 181.5 & 0,~0,~ 0,~  11.67 \\
200,~ 250 & 0.1 & -- & 0,~0,~ 50 & 0,~ 266.95 & 0,~0,~ 0,~  20.42 \\
200,~ 250 & 0.2 & -- & 0,~0,~ 50 & 0,~ 350.8 & 0,~0,~ 0,~  38.34 \\
\hline 
\hline 
400,~ 200 & 0.05 & -- & 0,~0,~ 50 & 0,~ 98.26 & 0,~0,~ 0,~  4.37 \\ 
400,~ 200 & 0.1 & -- & 0,~0,~ 50 & 0,~ 172.72 & 0,~0,~ 0,~  10.15 \\ 
400,~ 200 & 0.2 & -- & 0,~0,~ 50 & 0,~ 186.36 & 0,~0,~ 0,~  16.49 \\ 
\hline 
400,~ 300 & 0.05 & -- & 0,~0,~ 50 & 0,~ 277.72 & 0,~0,~ 0,~  21.77 \\ 
400,~ 300 & 0.1 & -- & 0,~0,~ 50 & 0,~ 400.8 & 0,~0,~ 0,~  47.98 \\ 
400,~ 300 & 0.2 & -- & 0,~0,~ 50 & 0,~ 595.98 & 0,~0,~ 0,~  98.65 \\ 
\hline 
400,~ 400 & 0.05 & -- & 0,~0,~ 50 & 0,~ 538.22 & 0,~0,~ 0,~  79.07 \\ 
400,~ 400 & 0.1 & -- & 0,~0,~ 50 & 0,~ 676.76 & 0,~0,~ 0,~  138.08 \\ 
400,~ 400 & 0.2 & -- & 0,~0,~ 50 & 0,~ 971.5 & 0,~9,~ 0,~  169.35 \\ 
\hline 
400,~ 450 & 0.05 & -- & 0,~0,~ 50 & 0,~ 745.28 & 0,~1,~ 0,~  106.77 \\ 
400,~ 450 & 0.1 & -- & 0,~0,~ 50 & 0,~ 802.2 & 0,~6,~ 0,~  160.33 \\ 
400,~ 450 & 0.2 & -- & 0,~0,~ 50 & 0,~ 1214.22 & 0,~18,~ 0,~  203.77 \\ 
\hline 
\hline 
500,~ 300 & 0.05 & -- & 0,~0,~ 50 & 0,~ 233.82 & 0,~0,~ 0,~  22.42 \\ 
500,~ 300 & 0.1 & -- & 0,~0,~ 50 & 0,~ 362.28 & 0,~0,~ 0,~  49.36 \\ 
500,~ 300 & 0.2 & -- & 0,~0,~ 50 & 0,~ 439 & 0,~0,~ 0,~  66.16 \\ 
\hline 
500,~ 400 & 0.05 & -- & 0,~0,~ 50 & 0,~ 568.55 & 0,~0,~ 0,~  74.59 \\ 
500,~ 400 & 0.1 & -- & 0,~0,~ 50 & 0,~ 651.22 & 0,~2,~ 0,~  124.53 \\ 
500,~ 400 & 0.2 & -- & 0,~0,~ 50 & 0,~ 787.58 & 0,~7,~ 0,~  163.85 \\ 
\hline 
500,~ 500 & 0.05 & -- & 0,~0,~ 50 & 0,~ 825.8 & 0,~3,~ 0,~  146.48 \\ 
500,~ 500 & 0.1 & -- & 0,~0,~ 50 & 0,~ 1234.92 & 0,~23,~ 0,~  189.37 \\ 
500,~ 500 & 0.2 & -- & 0,~0,~ 50 & 0,~ 1378.42 & 0,~30,~ 0,~  223.14 \\ 
\hline
\hline
1000,~ 500 & 0.05 & -- & 0,~0,~ 25 & 0,~ 609.24 & 0,~3,~ 0,~  150.16 \\ 
1000,~ 500 & 0.1 & -- & 0,~0,~ 25 & 0,~ 1199.32 & 0,~7,~ 0,~  204.81 \\ 
1000,~ 500 & 0.2 & -- & 0,~0,~ 25 & 0,~ 938.84 & 0,~12,~ 0,~  217.7 \\ 
\end{tabular}
\caption{Table for 200 through 1000 taxa representable by a persistent phylogeny. Galled-tree computations were not run on 
these data.}
\label{t4}
\end{center}
\end{table}

\section*{Acknowledgements} I would like to thank Anna Paola Carrieri for introducing me to persistent phylogeny,
and Paola Bonizzoni and Gianluca Della Vedova for helpful communications during the writing of this paper. Research 
partially supported by grants IIS-0803564, CCF-1017580, IIS-1219278 from the National Science Foundation.

%\bibliography{nconniebib}

\begin{thebibliography}{10}

\bibitem{BBDT12}
P.~Bonizzoni, C.~Braghin, R.~Dondi, and G.~Trucco.
\newblock The binary perfect phylogeny with persistent characters.
\newblock {\em Theoretical Computer Science}, 454:51--63, 2012.

\bibitem{BC2014arX}
P.~Bonizzoni, A.~P. Carrieri, G.~D. Vedova, and G.~Trucco.
\newblock Algorithms for the constrained perfect phylogeny with consistent
  characters, 2014.
\newblock arXiv:1405.7497v1.

\bibitem{BC2014}
P.~Bonizzoni, A.~P. Carrieri, G.~D. Vedova, and G.~Trucco.
\newblock Explaining evolution via constrained persistent perfect phylogeny.
\newblock {\em BMC Genomics}, 15(Suppl 6):S10, 2014.

\bibitem{BCDDP2013}
P.~Bonizzoni, A.P. Carrieri, G.~Della~Vedova, R.~Dondi, and T.M. Przytycka.
\newblock When and how the perfect phylogeny model explains evolution.
\newblock In N.~Jonoska and M.~Saito, editors, {\em Discrete and Topological
  Models in Molecular Biology}, Natural Computing Series, chapter~4. Springer,
  2013.

\bibitem{CG2012}
N.~C. Crawford and T.~C.~Glenn et~al.
\newblock More than 1000 ultraconserved elements provide evidence that turtles
  are the sister group of archosaurs.
\newblock {\em Biology Letters}, 8:783 --– 786, 2012.

\bibitem{Dollo}
L.~Dollo.
\newblock Le lois de l'\'evolution.
\newblock {\em Bulletin de la Societ\'e Belge de G\'eologie de Pal\'eontologie
  et d'Hydrologie}, 7:164--–167, 1893.

\bibitem{FEL04}
J.~Felsenstein.
\newblock {\em Inferring Phylogenies}.
\newblock Sinauer, 2004.

\bibitem{GUND2015}
G.~Gundem and S.~Bova et~al.
\newblock The evolutionary history of lethal metestatic prostate cancer.
\newblock {\em Nature}, 520:353–357, 2015.

\bibitem{GUSWEB}
D.~Gusfield.
\newblock wwwcsif.cs.ucdavis.edu/\verb|~|gusfield/.

\bibitem{GUS97}
D.~Gusfield.
\newblock {\em Algorithms on Strings, Trees and Sequences: Computer Science and
  Computational Biology}.
\newblock Cambridge University Press, 1997.

\bibitem{Recbook2014}
D.~Gusfield.
\newblock {\em ReCombinatorics: The Algorithmics of Ancestral Recombination
  Graphs and Explicit Phylogenetic Networks}.
\newblock MIT Press, 2014.

\bibitem{GELOPT04}
D.~Gusfield, S.~Eddhu, and C.~Langley.
\newblock Optimal, efficient reconstruction of phylogenetic networks with
  constrained recombination.
\newblock {\em Journal of Bioinformatics and Computational Biology},
  2(1):173--213, 2004.

\bibitem{GFB07}
D.~Gusfield, Y.~Frid, and D.~Brown.
\newblock Integer programming formulations and computations solving
  phylogenetic and population genetic problems with missing or genotypic data.
\newblock In {\em Proceedings of 13th Annual International Conference on
  Combinatorics and Computing}, pages 51--64. LNCS 4598, Springer, 2007.

\bibitem{Hem2010}
A.M. Heimberg and R.~Cowper-Sal lari~et al.
\newblock {MicroRNAs} reveal the interrelationships of hagfish, lampreys and
  gnathostomes and the nature of the ancestral vertebrate.
\newblock {\em Proceedings of the National Academy of Sciences (USA)},
  107:19379--19383, 2010.

\bibitem{Hillis1999}
D.~M. Hillis.
\newblock {SINEs} of the perfect character.
\newblock {\em Proceedings of the National Academy of Sciences (USA)},
  96:9979--9981, 1999.

\bibitem{HU2002}
R.~Hudson.
\newblock Generating samples under the {W}right-{F}isher neutral model of
  genetic variation.
\newblock {\em Bioinformatics}, 18(2):337--338, 2002.

\bibitem{PPSS}
I.~Pe'er, T.~Pupko, R.~Shamir, and R.~Sharan.
\newblock Incomplete directed perfect phylogeny.
\newblock {\em SIAM Journal on Computing}, 33:590--607, 2004.

\bibitem{Przyt07stable}
T.~Przytycka.
\newblock Stability of characters and construction of phylogenetic trees.
\newblock {\em Journal of Computational Biology}, 14:539--549, 2007.

\bibitem{Przyt07}
T.~Przytycka, G.~Davis, N.~Song, and D.~Durand.
\newblock Graph theoretical insights into evolution of multidomain proteins.
\newblock {\em Journal of Computational Biology}, 13:351--363, 2006.

\bibitem{RXSB2006}
D.~A. Ray, J.~Xing, A-H. Salem, and M.~A. Batzer.
\newblock {SINEs} of the {\it nearly} perfect character.
\newblock {\em Systematic Biology}, 55:928--935, 2006.

\bibitem{KooninDollo2006}
I.~B. Rogozin, Y.~I. Wolf, V.~N. Babenko, and E.~V. Koonin.
\newblock Dollo parsimony and the reconstruction of genome evolution.
\newblock In V.~A. Albert, editor, {\em Parsimony, Phylogeny, and Genomics}.
  Oxford University Press, 2006.

\bibitem{RH00}
A.~Rokas and P.~Holland.
\newblock Rare genomic changes as a tool for phylogenetics.
\newblock {\em Trends in Evolution and Ecology}, 15:454--459, 2000.

\bibitem{ZhengPrzyt07}
J.~Zheng, I.B. Rogozin, E.V. Koonin, and T.M. Przytycka.
\newblock Support for the {\it coelomata} clade of animals from a rigorous
  analysis of the pattern of intron conservation.
\newblock {\em Molecular Biology and Evolution}, 24:2583--2592, 2007.

\end{thebibliography}
%\input{ack}
\end{document}